\documentclass{ws-procs9x6}

\newcommand{\dels}{\delta q^s}
\newcommand{\delv}{\delta q^v}

\newcommand{\bii}{\scriptscriptstyle b_{1} }
\newcommand{\biNN}{\scriptscriptstyle b_{1}NN}
\newcommand{\hi}{\scriptscriptstyle h_{1}}
\newcommand{\hiNN}{\scriptscriptstyle h_{1}NN}

\newcommand{\nn}{\nonumber \\}

\begin{document}

\title{TRANSVERSITY VIA EXCLUSIVE $\pi$--ELECTROPRODUCTION}
\author{GARY R. GOLDSTEIN$^*$}

\address{Department of Physics and Astronomy, Tufts University,\\
Medford, MA 02155, USA \\
$^*$E-mail: gary.goldstein@tufts.edu}

\author{SIMONETTA LIUTI and SAEED AHMAD}

\address{Physics Department, University of Virginia,\\
Charlottesville, VA 22904, USA}

\begin{abstract}
Exclusive $\pi^0$ electroproduction from nucleons can be related to transversity, the tensor charge, and other quantities related to transversity. This process isolates C-parity odd and chiral odd combinations of t-channel exchange quantum numbers. In a hadronic picture the meson production amplitudes for intermediate energy and $Q^2$ are determined by C--odd Regge exchanges with final state interactions. In terms of partonic degrees of freedom, the helicity structure for this C-odd process relates to the quark helicity flip, or chiral odd generalized parton distributions (GPDs). Parameterization of chiral odd GPDs, using constraints of form factors, deep inelastic scattering and lattice results leads to predictions for various cross sections and asymmetries that will be sensitive to the transversity.
\end{abstract}

\keywords{Transversity; Regge; GPD; Electroproduction; $\pi^0$ production.}

\bodymatter

\section{Transversity questions}\label{intro:sec1}
What is Transversity? It is a concept introduced in the 1970's by Goldstein and Moravcsik~\cite{g_mor} as the appropriate spin quantization for two body scattering amplitudes, e.g. $f_{a,b;c,d}(s,t)$ that is most easily related to single spin asymmetry or azimuthal asymmetry observables. The transversity of a massive particle in its rest frame is its spin projection along an axis perpendicular to the two body scattering plane ($S\cdot p_{in} \times p_{out}$). In terms of helicities, transversity for spin  $\frac{1}{2}$ particles is $|\pm \frac{1}{2} )_T=\{|+\frac{1}{2}\rangle\pm (i)|-\frac{1}{2}\rangle\}/\sqrt{2}$.

The tensor charge is the first moment or the norm of the parton transversity distribution, $h_1(x)$. It is defined as the transversely polarized nucleon matrix element of local quark field operators,
$$
\langle P,S_T|\bar{\psi}\sigma^{\mu\nu}\gamma_5 \frac{\lambda^a}{2} \psi |P,S_T\rangle
=2\delta q^a (\mu^2) (P^\mu S_T^\nu- P^\nu S_T^\mu).$$
Like other charges, it is the integral of a distribution $(\delta q^a(x)-\delta \bar{q}^a(x))$, where  $\delta q^a(x)=h_1^a(x)$ is the {\bf transversity distribution}. It is essentially the probability to find a transversity $+\frac{1}{2}$ quark in a nucleon of transversity $+\frac{1}{2}$. Unlike the longitudinal distribution $g_1(x)$, $h_1(x)$ receives no contributions from gluons.

An important question is how the tensor charge can be determined, theoretically and experimentally? Some predictions and fits from various processes give: ($\delta u =1.26, \delta d = -0.17$) from QCD Sum rules (He and Ji,~\cite{heji}); ($\delta (u-d)= 1.09 ± 0.02$) from Lattice (QCDSF, M. Gockeler {\it et al.},~\cite{gock}); ($\delta u= 0.48\pm0.09, \delta d=-0.62\pm0.30$) from phenomenological analysis (Anselmino {\it et al.},~\cite{anselm}); ($\delta u= 0.58\pm0.20, \delta d =-0.11\pm0.20$) from axial vector dominance (Gamberg and Goldstein,~\cite{g_g}). 

The Gamberg and Goldstein theoretical model is based on axial vector dominance by the $b_1$(1235) and $h_1$(1170)--$h_1^\prime$(1380), with $J^{PC}=1^{+-}$, that couple to the tensor Dirac matrix $\sigma^{\mu\nu}\gamma_5$. The Dirac matrix has C-parity minus, which is a crucial fact. The resulting formulae for the isovector and isoscalar tensor charges are 
\begin{equation}
\delv =\frac{f_{\bii}g_{\biNN}\langle k_{\perp}^2\rangle }{\sqrt{2} M_N
M_{\bii}^2}\, ,\quad
\dels = \frac{ f_{\hi}g_{\hiNN}\langle k_{\perp}^2\rangle }{\sqrt{2} M_N
M_{\hi}^2},
\nn
\end{equation}
Because the axial vector couplings involve an additional angular momentum, to obtain the tensor structure a transverse momentum enters the coupling - the pure pole term decouples at zero momentum transfer. The interpretation that was adopted was that the coupling involves the quark constituents and thus does not vanish at zero momentum transfer. The average transverse momentum thus gives non-zero results. This led to three questions. How could the quarks be explicitly represented in the interaction with the axial vectors? The approach to answering this lay in the GPDs, particularly in the ERBL region. Hence exclusive processes should be considered, where the GPDs provide a description of hard scattering from the constituents. Secondly, how could the pole at the axial mass extrapolate to the $t=0$ limit, where the tensor charge is evaluated? This suggests the  Regge pole approach, which naturally allows extrapolation from physical poles to the physical scattering region, $t<0$, up to $t=t_{min}$, which approaches $t=0$ for asymptotic energies. Hence there is an interplay between a partonic description and a hadronic description of the tensor charge and transversity. What kind of reaction would single out the quantum numbers of the axial exchanges? For this, the exclusive photoproduction and electroproduction of $\pi^0$ or $\eta$ mesons from nucleons have C-parity odd in the t-channel and hence can accommodate the appropriate axial vector exchanges.

\begin{figure} 
\psfig{file=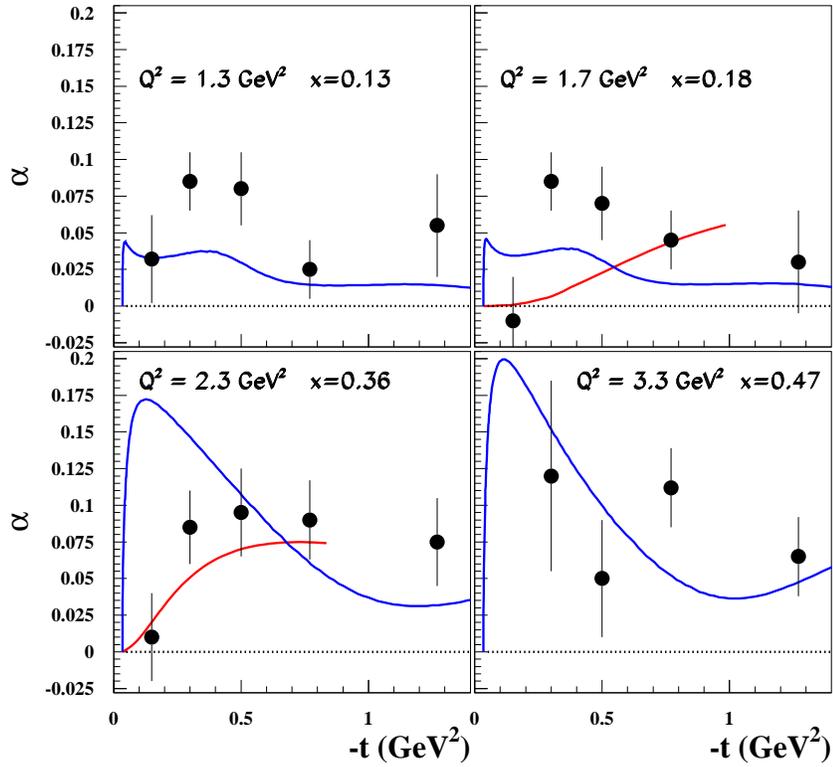,width=4.5in} 
\caption{Beam asymmetry in Regge and GPD pictures. Data from ref~\cite{demasi}.} 
\label{Fig:MV} 
\end{figure}

\section{Hadronic and partonic models}
We will consider the exclusive reactions, $e + p \rightarrow e^\prime + \pi^0 + p^\prime$ and the related $\eta$ and neutron target processes. The relevant subprocess is $\gamma^*+p\rightarrow \pi^0 + p^\prime$. The t-channel exchange picture involves C-parity odd, chiral odd states that include the $1^{+-}$ $b_1$ and $h_1$ mesons ($q+{\bar q}$; $S=0, L=1$ mesons) and the vector mesons, the $1^{--}$ $\rho^0$ and $\omega$ (($q+{\bar q}$; $S=1, L=0$ mesons). These axial vector mesons couple to the nucleon via the Dirac tensor $\sigma^{\mu\nu}\gamma_5$, while the vector mesons couple via $\gamma^\mu$ and/or $\sigma^{\mu\nu}$. Because of the C-parity there is no $\gamma^\mu \gamma^5$ coupling. This is quite significant in the GPD perspective - only chiral odd GPDs are involved, contrary to the accepted formulation~\cite{mank}. While ref.~\cite{mank} indicates that C-parity odd exchanges of 3 gluons, like the ``odderon", are allowed, the authors relate the process to chiral even GPDs that can involve $1^{++}$ exchange quantum numbers. This can be the case for charged pseudoscalar production, where there is not a C-parity eigenstate in the t-channel, but not for the neutral case, which has {\it definite odd C-parity}. Factorization issues for these processes have not been addressed for this C-parity odd case~\cite{CFS}, although vector production has received considerable attention~\cite{Diehl_Pire}

\begin{figure} 
\psfig{file=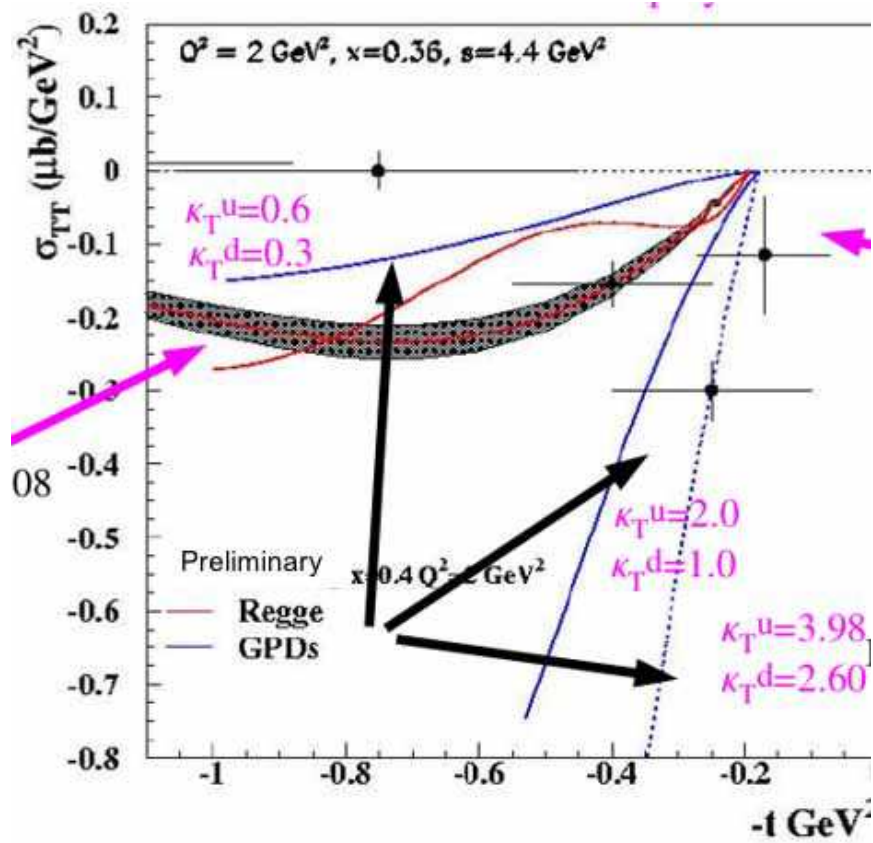,width=4.5in} 
\caption{$d\sigma_{TT}/dt$ Regge (wavy line) and GPD pictures (with 3 sets of $\kappa_T$ pairs). Data preliminary.} 
\label{Fig:TT} 
\end{figure}

\section{Regge cut model}
We proceed with the hadronic picture, the Regge model for $\pi^0$ electroproduction. A successful Regge cut model was developed to fit photoproduction data many years ago~\cite{g_o}. That model essentially involves as input the vector and axial vector meson trajectories that factorize into couplings to the on-shell $\gamma+\pi^0$ vertex and the nucleon vertex. The cuts or absorptive corrections destroy that factorization, but fill in the small $t$ and $t \approx -0.5$ amplitude zeroes. To connect to electroproduction, the upper vertex factor must acquire $Q^2$ dependence. This is accomplished by replacing the elementary, t-dependent couplings with $Q^2$ dependent transition form factors. In this Regge picture the factorization for the longitudinal virtual photon is not different from the transverse photon, except for the additional power of $Q^2$ for the longitudinal case. This is in contrast to the proofs of factorization for the longitudinal case in the GPD picture, while reaction initiation by transverse photons is not expected to factorize into a handbag picture~\cite{CFS}. With our form factor approach to the upper vertex (including Sudakov factors to soften the endpoint singularities) we anticipate  a similar factorization for the transverse case. 

The Regge picture is implemented by singling out the 6 independent helicity amplitudes and noting that at large $s$ and small $|t|$ the leading natural parity and unnatural parity Regge poles contribute to opposite sums and differences of pairs of helicity amplitudes. 

\section{Transversity, Chiral odd GPDs and Helicity}
Now the crucial connection to the 8 GPDs that enter the partonic  description of electroproduction is through the helicity decomposition~\cite{diehl}, where, for example, one of the chiral even helicity amplitudes is given by 

$A_{++,++}(X,\xi,t)=\frac{\sqrt{1-\xi^2}}{2}(H^q+{\tilde H}^q-\frac{\xi^2}{1-\xi^2}(E^q+{\tilde E}^q)),$

while one of the chiral odd amplitudes is given by

$A_{++,--}(X,\xi,t)=\sqrt{1-\xi^2}(H_T^q+\frac{t_0-t}{4M^2}{\tilde H}_T^q-\frac{\xi}{1-\xi^2}(\xi E_T^q+{\tilde E}_T^q)).$

There are relations to PDFs, $H^q(X,0,0)=f_1^q(X)$, ${\tilde H}^q(X,0,0)=g_1^q(X)$, $H_T^q(X,0,0)=h_1^q(X)$. The first moments of these are the charge, the axial charge and the tensor charge, for each flavor $q$, respectively. Further, the first moments of $E(X,0,0)$ and $2{\tilde H}_T^q(X,0,0)+E_T^q(X,0,0)$ are the anomalous moments $\kappa^q, \kappa_T^q$, with the latter defined by Burkardt~\cite{burk}.

Chiral even GPDs have been modeled in a thorough analysis~\cite{ahlt} , based on diquark spectators and Regge behavior at small $X$, and consistent with constraints from PDFs, form factors and lattice calculations. That analysis is used to obtain chiral odd GPDs via a multiplicative factor that fits the phenomenological $h_1(x)$~\cite{anselm}. With that {\it ansatz} the observables can be determined in parallel with the Regge predictions.

\section{Exclusive $\pi^0$ electroproduction observables}
The differential cross section for pion electroproduction off an unpolarized target is
\begin{equation}
\frac{d^4\sigma}{d\Omega dx d\phi dt} = \Gamma \left\{ \frac{d\sigma_T}{dt} + \epsilon_L \frac{d\sigma_L}{dt} + \epsilon \cos 2\phi \frac{d\sigma_{TT}}{dt} 
+ \sqrt{2\epsilon_L(\epsilon+1)} \cos \phi \frac{d\sigma_{LT}}{dt} \right\}.
\nn
\end{equation}
Each observable involves bilinear products of helicity amplitudes, or GPDs. For example, the cross section for the virtual photon linearly polarized out of the scattering plane minus that for the scattering plane is 
\begin{equation}
\frac{d\sigma_{TT}}{dt} = \mathcal{N} \, \frac{1}{s \mid P_{CM} \mid^2} %
  2 \Re e \left( f_{1,+;0,+}^*f_{1,-;0,-} - f_{1,+;0,-}^* f_{1,-;0,+} \right).
\nn
\end{equation}  
Another relevant observable is the traget transverse polarization asymmetry
\begin{equation}
\label{AUT}
A_{UT} =  \frac{2 \Im m (f_1^*f_3 - f_4^*f_2)}{{\displaystyle
\frac{d\sigma_T}{dt}}}.
\end{equation}

\section{GPD parameterization}
We performed calculations using a phenomenologically constrained 
model from the parametrization of Refs.\cite{ahlt}.
We summarize the AHLT model for the unpolarized GPD. The parameterization's form is:

\[ H(X,\zeta,t) = G(X,\zeta,t) R(X,\zeta,t), \]

\noindent
where $R(X,\zeta,t)$ is a Regge motivated term that 
describes the low $X$ and $t$ behaviors, while the contribution of 
$G(X,\zeta,t)$, obtained using a spectator model, is centered at intermediate/large values 
of $X$:
\begin{equation}
\label{diq_zeta}
G(X,\zeta,t) = 
{\cal N} \frac{X}{1-X} \int d^2{\bf k}_\perp \frac{\phi(k^2,\lambda)}{D(X,{\bf k}_\perp)}
\frac{\phi({k^{\prime \, 2},\lambda)}}{D(X,{ \bf k}_\perp^\prime)}. 
\nn  
\end{equation}
Here $k$ and $k^\prime$ are the initial and final quark momenta respectively; explicit expressions
are given in \cite{ahlt}. 
The 
$\zeta=0$ behavior is constrained by enforcing both the forward limit:
$H^q(X,0,0)  =  q_{val}(X)$, 
where $q_{val}(X)$ is the valence quarks distribution, and  
the following relations:
\begin{subequations}
\begin{eqnarray}
\int_0^1 dX H^q(X,\zeta,t)  =  F_1^q(t), \; \; \; \; 
\int_0^1 dX E^q(X,\zeta,t)  =  F_2^q(t),  
\end{eqnarray}
\label{FF}
\nn
\end{subequations}
which define the connection with the quark's contribution to the nucleon form factors.
Notice the AHLT parametrization does not make use of a
``profile function'' for the parton distributions, 
but the forward limit, $H(X,0,0) \equiv q(X)$, 
is enforced non trivially. This affords us the flexibility that 
is necessary to model the behavior at $\zeta, \, t \neq 0$. 
$\zeta$-dependent constraints are given by the higher moments of GPDs. 

The $n=1,2,3$ moments of the NS combinations: $H^{u-d} = H^u-H^d$, and $E^{u-d} = E^u-E^d$ 
are available
from lattice QCD \cite{haeg}, $n=1$ corresponding to the nucleon 
form factors. In a recent analysis a parametrization was devised that takes into 
account all of the above constraints. The parametrization gives an excellent description 
of recent Jefferson Lab data in the valence region.  

The connection to the transversity GPDs is carried out similarly to Refs.\cite{anselm} for the
forward case by setting:
\begin{equation}
H_T^q(X,\zeta,t)  =  \delta q H^{q, val}(X,\zeta,t) 
\nn
\end{equation}
\begin{equation}
\overline{E}_T^q  \equiv  2 \widetilde{H}_T + E_T  =  \kappa_T^q H_T^q(X,\zeta,t)  
\nn   
\end{equation}
where $\delta q$ is the tensor charge, and $\kappa_T^q$ is the tensor anomalous
moment  introduced, and connected to the transverse component of the total angular momentum
in \cite{burk}. 
Notice that our unpolarized GPD model can be adequately extended to describe $H_T$ 
since it was developed in the valence region, and transversity involves valence quarks only.

\begin{figure} 
\psfig{file=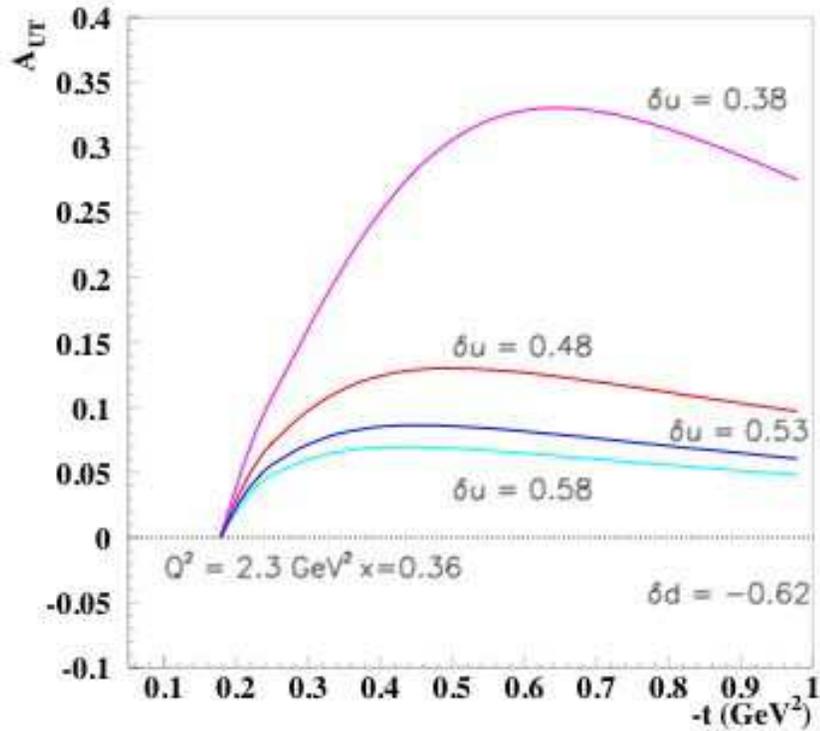,width=4.5in} 
\caption{Transverse spin asymmetry, $A_{UT}$, Eq.(\ref{AUT}).} 
\label{Fig:AUT} 
\end{figure}

In Fig.\ref{Fig:AUT} we show the sensitivity of $A_{UT}$ to to the values of the u-quark 
and d-quark tensor charges. The values in the figure 
were taken by varying up to $20 \%$ the values of the tensor charge extracted from the global 
analysis of Ref.\cite{anselm}, {\it i.e.}  $\delta u=0.48$ and $\delta d = -0.62$, and fixing the transverse
anomalous magnetic moment values to
$\kappa_T^u = 0.6$ and $\kappa_T^d = 0.3$.
This is the main result of this contribution: it summarizes our 
proposed method for a practical extraction of 
the tensor charge from $\pi^o$ electroproduction experiments. 
Therefore our model can be used to constrain the range of values allowed by the data~\cite{agl}.
\section*{Acknowledgments}
This work is supported by the U.S. Department of Energy grants no. DE-FG02-92ER40702 (G.R.G.) and no. DE-FG02-01ER4120 (S.L). We thank L. Gamberg for initial conversations.




\end{document}